# A theoretical model of Dark Energy Stars in Einstein-Gauss-Bonnet Gravity

Manuel Malaver[1,2], Hamed Daei Kasmaei[3], Rajan Iyer[4], Shouvik Sadhukhan[5], Alokananda Kar[6]

[1]Bijective Physics Institute, Idrija, Slovenia

[2]Maritime University of the Caribbean, Department of Basic Sciences, Catia la Mar, Venezuela.
 Email: **mmf.umc@gmail.com**

[3]Islamic Azad University, Department of Applied Mathematics and Computer Science, Central Tehran Branch, Tehran, Iran
 Email: **hamedelectroj@gmail.com**

[4]Environmental Materials Theoretical Physicist, Department of Physical Mathematics Sciences Engineering
  Project Technologies, Engineeringinc International Operational Teknet Earth Global, Tempe, Arizona, United
  States of America
  Email: engginc@msn.com

[5]Indian Institute of Technology Kharagpur, Department of Physics & Meteorology, Kharagpur, India
  Email: shouvikphysics1996@gmail.com

[6]University of Calcutta, Department of Physics, Kolkata West Bengal, India
  Email: alokanandakar@gmail.com

**Abstract:** Dark energy stars research is an issue of great interest since recent astronomical observations with respect to measurements in distant supernovas, cosmic microwave background and weak gravitational lensing confirm that the universe is undergoing a phase of accelerated expansion and this cosmological behavior is caused by the presence of a cosmic fluid which has a strong negative pressure that allows to explain the expanding universe. In this paper, we obtained new relativistic stellar configurations within the framework of Einstein-Gauss-Bonnet (EGB) gravity considering negative anisotropic pressures and the equation of state $p_r = \omega \rho$ where $p_r$ is the radial pressure, $\omega$ is the dark energy parameter, and $\rho$ is the dark energy density. We have chosen a modified version of metric potential proposed by Korkina-Orlyanskii (1991). For the new solutions we checked that the radial pressure, metric coefficients, energy density and anisotropy are well defined and are regular in the interior of the star and are dependent of the values of the Gauss-Bonnet coupling constant. The solutions found can be used in the development of dark energy stars models satisfying all physical acceptability conditions, but the causality condition and strong energy condition cannot be satisfied.

**Keywords:** Dark energy stars, EGB gravity, Metric potential, Coupling constant, Causality condition, Strong energy condition.

# 1. Introduction

The origin of the energy content in the universe is a fundamental issue in cosmology and currently there is sufficient observational evidence as measurements of supernovas of type Ia and microwave background radiation that favors an accelerate expansion [1]. The explanation for this cosmological behavior in the framework of general relativity requires assuming that a considerable part of the universe consists of a hypothetical dark energy with a negative pressure component [2]. Many authors have suggested that this dark energy is a cosmic fluid parameterized by an equation of state $\omega = p/\rho < -1/3$ where p is the spatially homogeneous pressure and $\rho$ is the dark energy density [1-5]. Redshifts of distant supernovae SN-Ia measurements have made astrophysicists to surmise universe as showing an accelerated expansion with a coasting universe reminiscent of a vanishing cosmic mass density [6], while recent Hubble tension measurements have posited dynamical dark energy [7]. The range for which $\omega < -1$ has been denoted as phantom energy and possesses peculiar properties, such as negative temperatures and the energy density which increases to infinity in a finite time, resulting in a big rip [2,3,4]. It also provides a natural scenario for the existence of exotic geometries such as wormholes [5,8,9]. The upper limit of the mass of an ideal white dwarf has been astrophysically derived to be 0.91 solar mass [10]. Positron spectrum for 106 solar mass dark energy star signature heralds existence of dark energy stars undergoing seemingly nucleon decay, measurable in the form of positron emissions, since quarks and gluons inside nucleons have energies correlating to observed excess energies over typically calculated limit for both collapsed stars and the compact objects; at quantum critical region, of the center of galaxies, they seemingly morph into heavy particles within superfluids [11]. Einstein field equations have been solved choosing a particular Lorentzian-type density distribution function with the standard 4-dimensional spacetime as well as higher dimensions: 5D, 6D, and 11D; their results show that matter-energy density, radial as well as transverse pressures, anisotropy, and other quantities, especially the surface redshift seemed to suggest compact star 4U 1820-30 to be a strange quark star [12]. Model of a statically charged anisotropic fluid sphere in the Einstein Maxwell Gauss Bonnet (EMGB) theory of gravitation satisfies all the non-negativity nature of pressure and density, energy conditions, among other elementary physical requirements [13]. Massive star like a neutron star or a pulsar has been modelled [14] and an exact Maxwell-Einstein metric for a spherically symmetric static perfect fluid with mass and charge (Q) electrifying Tolman-VII metric, while meeting applicable physical boundary conditions satisfactorily wholly has been derived appropriately [15].

The notion of dark energy is that of a homogeneously distributed cosmic fluid and when extended to inhomogeneous spherically symmetric spacetimes, the pressure appearing in the equation of state is now a negative radial pressure, and the tangential pressure is then determined via the field equations [2,3]. Lobo [3] explored several configurations, by imposing specific choices for the mass function and studied the dynamical stability of these models by applying the general stability formalism developed by Lobo and Crawford [16]. Chan et al. [17] proposed that the mass function is a natural consequence of the Einstein´s field equations and considered a core with a homogeneous energy density, described by the Lobo´s first solution [3]. Malaver and Esculpi [18] presented a new model of dark energy star by imposing specific choice for the mass function that corresponds to an increase in energy density inside of the star. Bibi et al. [4] obtained a new class of solutions of the Einstein-Maxwell field equations which represents a model for dark energy stars with the equation of state $p_r=-\rho$ . Malaver et al. [19] found a new family of solutions to the Einstein-Maxwell system considering a particular form of the gravitational potential $Z(x)$ and the electric field intensity with a linear equation of state that represents a model of dark energy star. According to Chan et al. [17], the denomination dark energy is applied to fluids which violate only the strong energy condition given by $\rho+p_r+2p_t \geq 0$ where $\rho$ is the energy density, $p_r$ and $p_t$ are the radial pressure and tangential pressure, respectively.

The analysis of compact objects with anisotropic matter distribution is very important, because that the anisotropy plays a significant role in the studies of relativistic spheres of fluid [20-32]. Anisotropy is defined as $\Delta = p_t - p_r$ where $p_r$ is the radial pressure and $p_t$ is the tangential pressure. The existence of solid core, presence of type 3A superfluid [33], magnetic field, phase transitions, a pion condensation and electric field [34] are most important reasonable facts that explain the presence of tangential pressures within a star. Many astrophysical objects as X-ray pulsar, Her X-1, 4U1820-30 and SAXJ1804.4-3658 have anisotropic pressures. Bowers and Liang [32] include in the equation of hydrostatic equilibrium the case of local anisotropy. Bhar et al. [35] have studied the behavior of relativistic objects with locally anisotropic matter distribution considering the Tolman VII form for the gravitational potential with a linear relation between the energy density and the radial pressure. Malaver [36-37], Feroze and Siddiqui [38,39] and Sunzu et al.[40] obtained solutions of the Einstein-Maxwell field equations for charged spherically symmetric space-time by assuming anisotropic pressure.

Mathematical modeling within the framework of the general theory of relativity has been used to explain the behavior and structure of massive objects as neutron stars, quasars, black holes, pulsars and white dwarfs [41,42] and requires finding the exact solutions of the Einstein-Maxwell system [5]. A detailed and systematic analysis was carried out by Delgaty and Lake [43] which obtained several analytical solutions that can describe realistic stellar configurations.

Recently, astronomical observations of compact objects have allowed new findings of neutron stars and strange stars that adjust to the exact solutions of the 4-D Einstein field equations and the data on mass maximum, redshift and luminosity are some of the most relevant characteristics for verifying the physical requirements of these models [13]. A great number of exact models from the Einstein-Maxwell field equations have been generated by Gupta and Maurya [14], Kiess [15], Mafa Takisa and Maharaj [44], Malaver and Kasmaei [45], Malaver [46,47], Ivanov [48] and Sunzu et al [40]. In the development of these models, several forms of equations of state can be considered [49]. Komathiraj and Maharaj [50], Malaver [51], Bombaci [52], Thirukkanesh and Maharaj [53], Dey et al. [54] and Usov [34] assume linear equation of state for quark stars. Feroze and Siddiqui [38] considered a quadratic equation of state for the matter distribution and particular forms for the gravitational potential and electric field intensity. MafaTakisa and Maharaj [44] obtained new exact solutions to the Einstein-Maxwell system of equations with a polytropic equation of state. Thirukkanesh and Ragel [55] have obtained particular models of anisotropic fluids with polytropic equation of state which are consistent with the reported experimental observations. Malaver [56] generated new exact solutions to the Einstein-Maxwell system considering Van der Waals modified equation of state with polytropic exponent. Tello-Ortiz et al. [57] found an anisotropic fluid sphere solution of the Einstein-Maxwell field equations with a modified version of the Chaplygin equation of state.

The modified gravity helps in the resolution of several cosmological problems like negative pressure problems, late time acceleration problem, Dark energy [58] and exotic matter problems etc. Dark energy resolved those problems with alternative gravity technique but modified gravity is like an alternative to the dark energy models. Modification of gravity means the modification in its action. In general, the Einstein action contained with the first order function of Ricci scalar. This action provides simple expanding geometry but it is inefficient to give the definition of negative pressure or the cause behind the accelerated expansion. On the other-hand, Raychaudhuri equation discussed the conditions behind the accelerated expansion which also remain unsolved with first ordered Einstein action. That's why the multiple order has been brought in that action w.r.t. the Ricci scalar. The modified gravity also provides some similar result with Van-Der-Waals non-linear fluid system [59].

The higher order terms in Einstein action [60] is brought with the multiple powers of Ricci scalar. Those process to get the Lagrangian for the modified action is called Lovelock Lagrangian construction. In this paper we have used the order 2 modification where the Lagrangian has been constructed with Ricci scalar, Ricci tensor and Riemannian curvature tensor. It is also well known that the Gauss–Bonnet term features in heterotic string theory and the coupling constant carries the meaning of the string tension in that area. This is also a good reason to probe the behaviour of higher curvature invariants in gravitational field theory in view of the long time project to merge quantum field theory, of which string theory is a leading candidate, and gravitational physics.

The behavior and dynamics of the gravitational field can be extended to higher dimensions [61]. The history of higher dimensions goes back to the work done by Kaluza [62] and Klein [63] who introduced the concept of extra dimensions in addition to the usual four dimensions (4-D) to unify gravitational and electromagnetic interactions. Šorli and Čelan [64] developed a model of the time-invariant n-dimensional complex superfluid quantum space that presents a new solution to the unification of the forces. In general theory of relativity, the results obtained in four dimensions can be generalized in higher dimensional contexts and study the effects due to incorporation of extra space-time dimensions [65]. Within this framework, a very useful and fruitful generalization is the Einstein-Gauss-Bonnet gravity, which has generated a lot of interest among researchers and has been influenced by many scientists that are working in this field [66]. The modeling of compact objects in EGB gravity has shown that some physical variables are modified when they are compared to their 4-D counterparts, but the condition of the Schwarzschild constant density sphere has been demonstrated in EGB gravity [13]. Recently, Bhar et al. [67] performed a comparative study of compact objects in five dimensions (5-D) between EGB gravity and classical general relativity theory and found that many features as stability, causality and energy conditions remain unaffected in the stellar interior.

The aim of this paper is to generate new class of solutions which represents a potential model of dark energy stars whose equation of state is $p_r = \omega \rho$ and with anisotropic matter distribution, within the framework of Einstein-Gauss-Bonnet (EGB) gravity. We have used a modified form of gravitational potential proposed by Korkina-Orlyanskii [52]. The system of field equations has been solved to obtain analytic solutions which are physically acceptable. The paper is organized as follows: In Section.2, we present the framework of EGB gravity. The modified Einstein-Maxwell field equations with the Gauss-Bonnet coupling constant are presented in Section.3. With the chosen metric potential, we generate some new models for dark energy star within EGB gravity in Section.4. In Section.5, physical requirements for the new models are described. In Section.6, a physical analysis of the new solutions is performed. In final Section, we conclude.

## 2. Einstein-Gauss-Bonnet Gravity

The Gauss-Bonnet action in a $n$-dimensional space-time can be written as

$$S = \int \sqrt{-g} \left[ \frac{1}{2k_n^2} (R + \alpha L_{GB}) \right] d^n x + S_{matter} \tag{1}$$

where $\alpha$ is the Gauss-Bonnet coupling constant. The strength of the action $L_{GB}$ lies in the fact that despite the Lagrangian being quadratic in the Ricci tensor, Ricci scalar and the Riemann tensor, the equations of motion turn out to be second order quasi-linear which are compatible with Einstein's theory of gravity [66,67].

The EGB field equations may be written as

$$G_{ab} + \alpha H_{ab} = T_{ab} \tag{2}$$

where $G_{ab}$ represents the Einstein tensor, $T_{ab}$ is the total energy-momentum tensor and the Lanczos tensor $H_{ab}$ is given by

$$H_{ab} = 2(RR_{ab} - 2R_{ac}R_b^c - 2R^{cd}R_{acbd} + R_a^{cde}R_{bcde}) - \frac{1}{2} g_{ab} L_{GB} \tag{3}$$

where the Lovelock term has the form

$$L_{GB} = R^2 + R_{abcd}R^{abcd} - 4R_{cd}R^{cd} \tag{4}$$

## 3. Modified Einstein-Maxwell Field Equations with Gauss-Bonnet coupling constant

By taking $n=5$, the 5-dimensional line element for a static spherically symmetric space-time has the form

$$ds^2 = -e^{2\nu(r)}dt^2 + e^{2\lambda(r)}dr^2 + r^2(d\theta^2 + \sin^2\theta d\phi^2 + \sin^2\theta \sin^2\phi d\psi^2) \quad (5)$$

where the metric functions $e^\nu$ and $e^\lambda$ are the gravitational potentials. By considering the commoving fluid velocity as $u^a = e^{-\nu}\delta_0^a$, the EGB field equations (2) reduce to

$$\rho = \frac{3}{e^{4\lambda}r^3}\left(4\alpha\lambda' + re^{2\lambda} - re^{4\lambda} - r^2 e^{2\lambda}\lambda' - 4\alpha e^{2\lambda}\lambda'\right) \quad (6)$$

$$p_r = \frac{3}{e^{4\lambda}r^3}\left(-re^{4\lambda} + \left(r^2\nu' + r + 4\alpha\nu'\right)e^{2\lambda} - 4\alpha\nu'\right) \quad (7)$$

$$p_t = \frac{1}{e^{4\lambda}r^2}\left(-e^{4\lambda} - 4\alpha\nu'' + 12\alpha\nu'\lambda' - 4\alpha(\nu')^2\right) +$$
$$\frac{1}{e^{2\lambda}r^2}\left(1 - r^2\nu'\lambda' + 2r\nu' - 2r\lambda' + r^2(\nu')^2\right) +$$
$$\frac{1}{e^{2\lambda}r^2}\left(r^2\nu'' - 4\alpha\nu'\lambda' + 4\alpha(\nu')^2 + 4\alpha\nu''\right) \quad (8)$$

Here primes means a derivation with respect to the radial coordinates $r$ and $\rho$ is the energy density, $p_r$ is the radial pressure and $p_t$ is the tangential pressure. With the transformations $x = cr^2$, $Z(x) = e^{-2\lambda}$ and $y^2(x) = e^{2\nu}$ suggested by Durgapal and Bannerji [68] and with $c>0$ as arbitrary constant, the field equations (6)-(8) can be written as follows

$$\frac{\rho}{c} = -3\dot{Z} - \frac{3(Z-1)(1-4\beta\dot{Z})}{x} \quad (9)$$

$$\frac{p_r}{c} = \frac{3(Z-1)}{x} + \frac{6Z\dot{y}}{y} - \frac{6\beta(Z-1)Z\dot{y}}{xy} \quad (10)$$

$$\frac{p_t}{c} = 4Z\left[\beta(1-Z) + x\right]\frac{\ddot{y}}{y} + \left[\frac{2\beta Z(1-Z)}{x} - 2(x+\beta)\dot{Z} + 6Z - 2\beta Z\dot{Z}\right]\frac{\dot{y}}{y} + 2\left[\frac{Z-1}{x} + \dot{Z}\right] \quad (11)$$

where $\beta = 4\alpha c$ contains the Gauss-Bonnet coupling constant $\alpha$ and dots denote differentiation with respect to $x$.

In this paper, we assume the following equation of state

$$p_r = \omega\rho \tag{12}$$

where $\omega$ is the dark energy parameter and $-1 \leq \omega \leq -1/3$.

## 4. Generating novel Models of Dark Energy Stars in EGB gravity

Following Singh et al. [69], in this paper we made the choice for $Z(x)$ as

$$Z(x) = \frac{ax+1}{2ax+1} \tag{13}$$

where a is a real constant. This potential is regular at the stellar center and well behaved in the interior of the sphere. Substituting $Z(x)$ in equation.(9), we obtain

$$\rho = \frac{6aC(2a^2x^2 + 3ax + 1 + 2a\beta)}{(1+2ax)^3} \tag{14}$$

Replacing equation.(14) in equation.(12), we have for the radial pressure

$$p_r = \frac{\omega 6aC(2a^2x^2 + 3ax + 1 + 2a\beta)}{(1+2ax)^3} \tag{15}$$

and with $Z(x)$ and equation.(15), we have

$$\frac{\dot{y}}{y} = \frac{\omega[12a^3x^2 + 18a^2x + 12a^2\beta + 6a]}{(1+2ax)[6(ax+1)(2ax+1) + 6a\beta(ax+1)]} + \frac{3a(1+2ax)}{6(ax+1)(2ax+1) + 6a\beta(ax+1)} \tag{16}$$

Integrating (16) we obtain

$$y(x) = c_1(2ax+1)^A (ax+1)^B (a\beta + 2ax+1)^C \tag{17}$$

where the constants A, B and C are given by

$$A = 2\omega \tag{18}$$

$$B = -\frac{4a\beta\omega+1}{2(a\beta+1)} \tag{19}$$

$$C = \frac{a\beta(\omega+1)+3\omega}{2(a\beta-1)} \tag{20}$$

and $c_1$ is the constant of integration.

For the metric functions $e^{2\lambda}$ and $e^{2\nu}$ we have

$$e^{2\lambda} = \frac{2ax+1}{ax+1} \tag{21}$$

$$e^{2\nu} = c_1^2 (2ax+1)^{2A} (ax+1)^{2B} (a\beta+2ax+1)^{2C} \tag{22}$$

and the anisotropy can be written as

$$\Delta = \frac{4xc(ax+1)(a\beta+2ax+1)}{(2ax+1)^2} \left[ \frac{4(A^2-A)a^2}{(2ax+1)^2} + \frac{4ABa^2}{(2ax+1)(ax+1)} + \frac{8ACa^2}{(2ax+1)(a\beta+2ax+1)} + \frac{(B^2-B)a^2}{(ax+1)^2} + \frac{4BCa^2}{(ax+1)(a\beta+2ax+1)} + \frac{4a^2(C^2-C)}{(a\beta+2ax+1)^2} \right]$$
$$+ \frac{2a\left[2a(1-2a\beta)x^2+(1-3a\beta)x\right]\left[\frac{2aA}{2ax+1}+\frac{aB}{ax+1}+\frac{2aC}{a\beta+2ax+1}\right]}{(1+2ax)^3} + \frac{2a^2x}{(1+2ax)^2}$$

$$(23)$$

## 5. Requirements of Physical Acceptability in EGB Gravity

For a model to be physically acceptable in EGB gravity, the following conditions should be satisfied [66,67]:

(i) The metric potentials $e^{2\lambda}$ and $e^{2\nu}$ assume finite values throughout the stellar interior and are singularity-free at the center $r=0$.

(ii) The energy density $\rho$ should be positive and a decreasing function inside the star, $\frac{d\rho}{dr} \leq 0$ in EGB gravity.

(iii) The radial pressure also should be positive and a decreasing function of radial parameter but for negative pressure, this condition is not satisfied.

(iv) The anisotropy is zero at the center $r=0$, i.e. $\Delta(r=0) =0$.

.

(v) Any physically acceptable model must satisfy the causality condition, that is, for the radial sound speed $v_{sr}^2 = \frac{dp_r}{d\rho}$ ,we should have $0 \leq v_{sr}^2 \leq 1$ but the dark energy case this condition nor is it satisfied.

(vi) The boundary of the star defined by $r=R$ should be matched with the Einstein –Gauss-Bonnet- Schwarzschild exterior solution given by

$$ds^2 = -F(r)dt^2 + \frac{dr^2}{F(r)} + r^2(d\theta^2 + \sin^2\theta d\phi^2 + \sin^2\theta \sin^2\phi d\psi^2)\tag{24}$$

where $R$ is the radius of the star and

$$F(r) = 1 + \frac{r^2}{4\alpha}\left(1 - \sqrt{1 + \frac{8M\alpha}{r^4}}\right)\tag{25}$$

In Equation. (25) ,$M$ is associated with the gravitational mass of the hypersphere.

## 6. Physical Analysis of the New Models

We present the physical analysis for the proposed new models:

The metric potentials $e^{2\lambda}$ and $e^{2\nu}$ have finite values and remain positive throughout the stellar interior. At the center, we have $e^{2\lambda(0)} = 1$ and $e^{2\nu(0)} = c_1^2(a\beta+1)^{2C}$. We show that in $r=0$, $\left(e^{2\lambda(r)}\right)'_{r=0} = \left(e^{2\nu(r)}\right)'_{r=0} = 0$ and this makes this result that is possible to verify that the gravitational potentials are regular at the center.

The energy density is positive and well behaved in the stellar interior. In the center, we have $\rho(r=0) = ac(1+2a\beta)$ and $p_r(r=0) = 6\omega ac(1+2a\beta)$, therefore the energy density will be non-negative in $r=0$ and $p_r(r=0) < 0$.

For the density gradient inside the stellar interior, we obtain

$$\frac{d\rho}{dr} = \frac{6ac(8a^2c^2r^3 + 6acr)}{(1+2acr^2)^3} - \frac{72a^2c^2(2a^2c^2r^4 + 3acr^2 + 8a\alpha c + 1)r}{(1+2acr^2)^4}\tag{26}$$

Using the first fundamental form that consist in the continuity of the metric functions and their derivatives across the boundary $r=R$, we have

$$\frac{2acR^2+1}{acR^2+1}=\frac{1}{1+\frac{R^2}{4\alpha}\left(1-\sqrt{1+\frac{8\alpha M}{R^4}}\right)} \quad (27)$$

$$c_1^2\left(2acR^2+1\right)^{2A}\left(acR^2+1\right)^{2B}\left(a\beta+2acR^2+1\right)^{2C}=1+\frac{R^2}{4\alpha}\left(1-\sqrt{1+\frac{8\alpha M}{R^4}}\right) \quad (28)$$

$$4\left[\frac{2AacR}{2acR^2+1}+\frac{BacR}{1+acR^2}+\frac{2CacR}{2acR^2+a\beta+1}\right]c_1^2\left(2acR^2+1\right)^{2A}\left(1+acR^2\right)^{2B}\left(2acR^2+a\beta+1\right)^{2C}$$

$$=-\frac{1}{2\alpha}\left[\frac{1-\sqrt{1+\frac{8\alpha M}{R^4}}}{\sqrt{1+\frac{8\alpha M}{R^4}}}\right]$$

(29)

and from the second fundamental form, we have

$$12a^3c^3R^4+18a^2c^2R^2+6ac+12a^2\beta c=0 \quad (30)$$

The equations (27-30) are the conditions that allow determining the parameters a, A, B, C that describe the model.

In the figures 1, 2 and 3 are represented the dependence of $\rho$, $\frac{d\rho}{dr}$ and $e^{2\lambda}$ with the radial coordinate for different values of coupling constant. In all the cases were investigated, it has been considered $a=0.0064$, $c=1$, $\omega=-1$ and the radius $R= 9\ Km$.

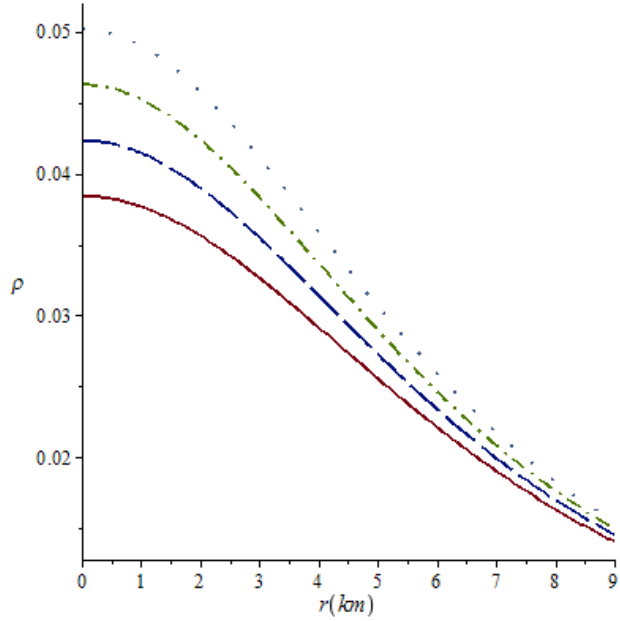

**Figure 1**. *Energy density against radial coordinate for α=0 (solid line); α=2 (long-dash line): α=4 (dashdot line); α=6 (spacedot line). In three cases a=0.0064, ω=-1 and c=1.*

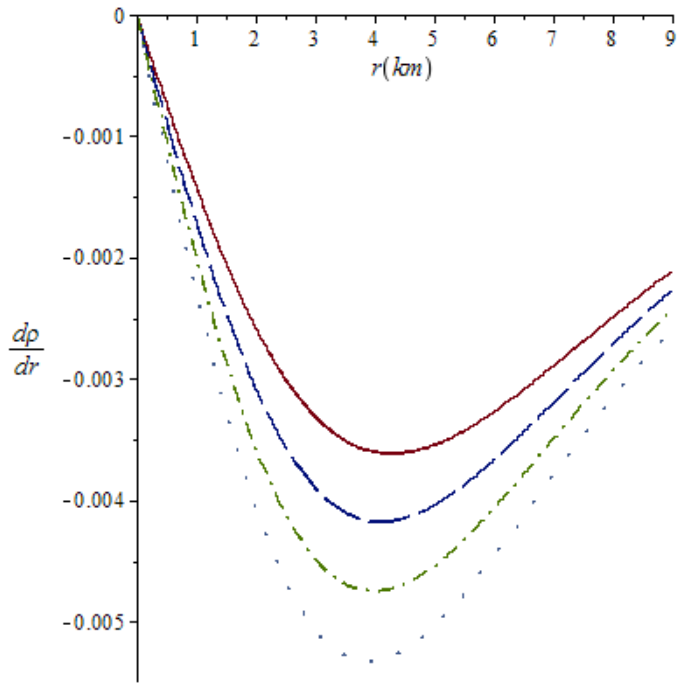

**Figure 2**. *Density gradient against radial coordinate for α=0 (solid line); α=2 (long-dash line): α=4 (dashdot line); α=6 (spacedot line). In all the cases a=0.0064, ω=-1 and c=1.*

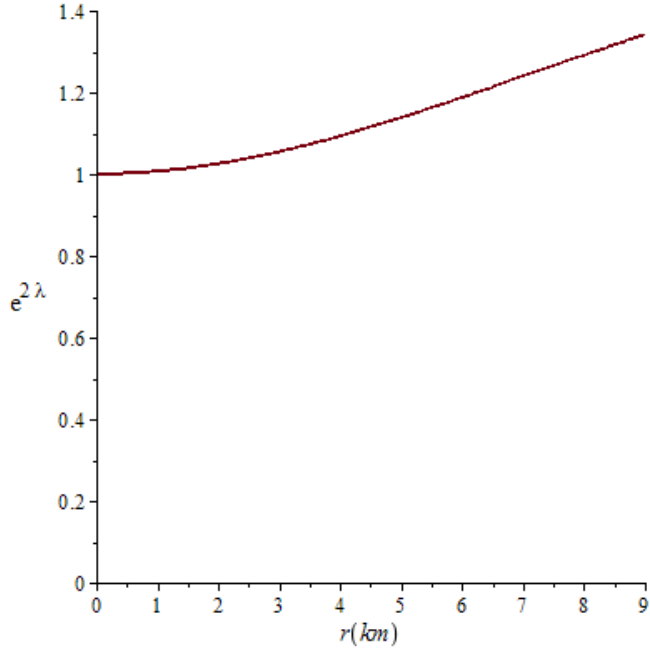

**Figure 3.** *Metric potential $e^{2\lambda}$ against radial coordinate for a=0.0064 and c=1.*

For different values of coupling constant, the energy density remains positive, continuous and is monotonically decreasing function throughout the stellar interior as noted in the Figure 1. It is also seen that the density increases with increasing α. The radial variation of energy density gradient has been shown in Figure 2, in which it is observed that $\frac{d\rho}{dr} < 0$ in EGB gravity and the metric potential $e^{2\lambda}$ in Figure 3 is a continuously growing function inside the star.

The Figures 4,5, 6 and 7 show the dependence of $e^{2\nu}$, $p_r$, anisotropy ∆ and strong energy condition (SEC) respectively with the radial parameter for the different values of coupling constant α. In all the cases, it has been considered *R= 9 Km, a=0.0064, c=1* and *ω = -1*.

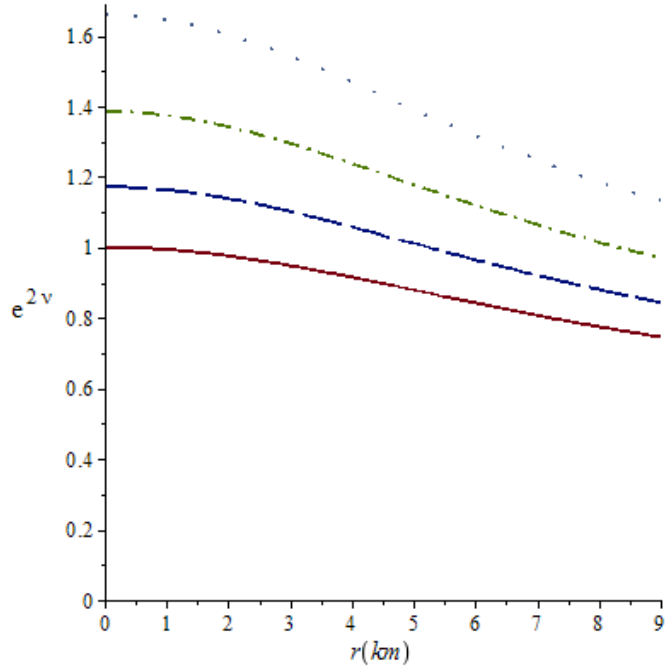

**Figure 4**. Metric potential $e^{2\nu}$ against radial coordinate for α=0 (solid line); α=2 (long-dash line): α=4 (dashdot line); α=6 (spacedot line). In all the cases a=0.0064, ω=-1 and c=1.

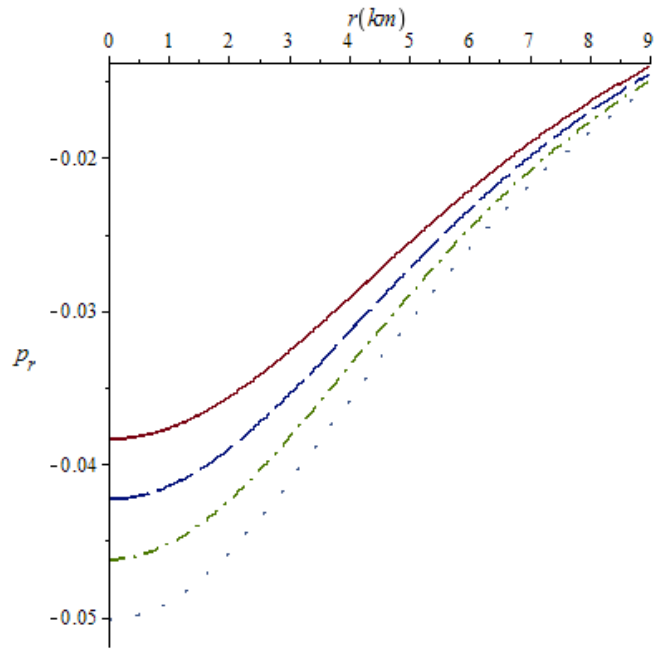

**Figure 5**. Radial pressure against radial coordinate for α=0 (solid line); α=2 (long-dash line): α=4 (dashdot line); α=6 (spacedot line). In all the cases a=0.0064, ω=-1 and c=1.

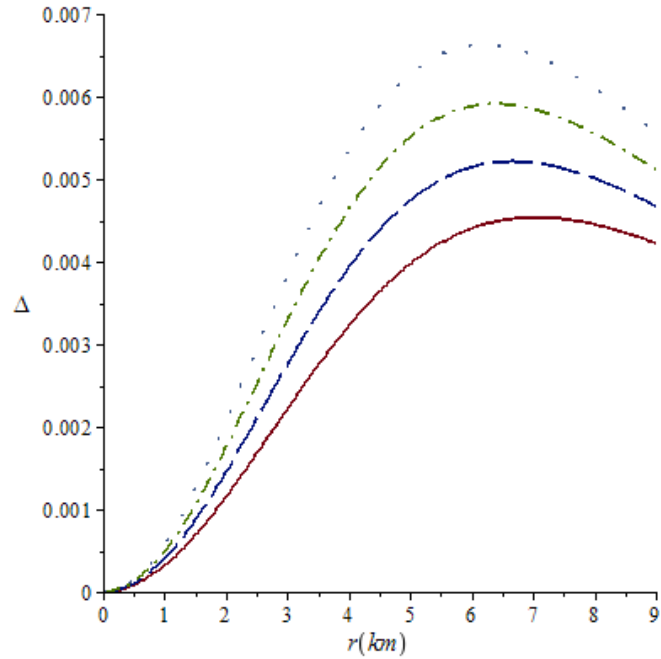

*Figure 6*. Anisotropy against radial coordinate for α=0 (solid line); α=2 (long-dash line): α=4 (dashdot line); α=6 (spacedot line). In all the cases a=0.0064, ω=-1 and c=1.

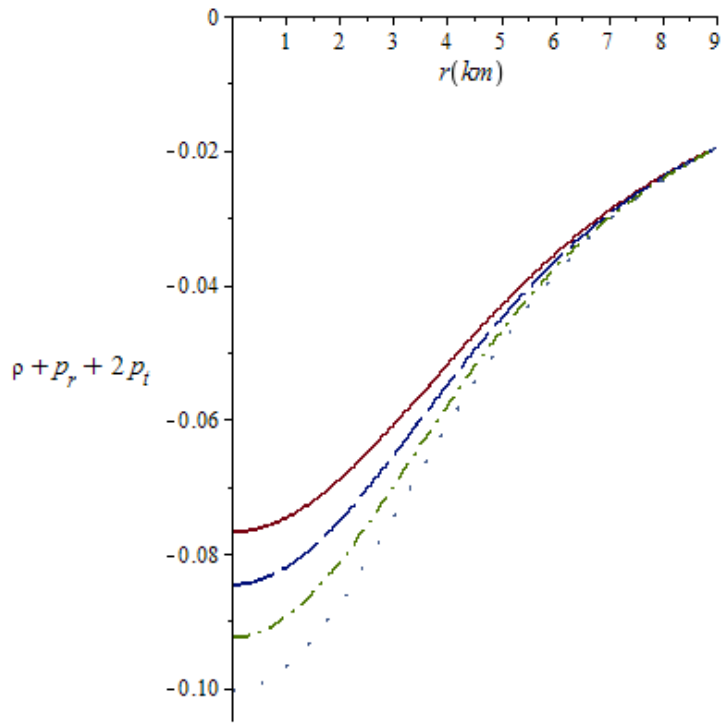

*Figure 7*. Strong energy condition $\rho+p_r+2p_t \geq 0$ against radial coordinate for α=0 (solid line); α=2 (long-dash line): α=4 (dashdot line); α=6 (spacedot line). In all the cases a=0.0064, ω=-1 and c=1.

In figure 4, the metric potential $e^{2\nu}$ is continuous, well behaved and increases with an increase of the values of α and from the equation.(15), the radial pressure is negative and not a decreasing function of the radial parameter ,but takes lower values when α is increased as shown in Figure 5. The anisotropic factor is plotted in Figure 6 and it shows that vanishes at the centre of the star, i.e. Δ(r=0) =0. We can also note that Δ admits higher values with a growth of α. The Figure 7 shows that the strong energy condition is violated for all α values considered when ω=-1.

In the figures 8, 9 and 10 have been represented the variation of $e^{2\nu}$, Δ and SEC with the radial coordinate for the different values of the dark energy parameter ω. In these cases, we have considered α=2 and R= 9 Km, a=0.0064, c=1.

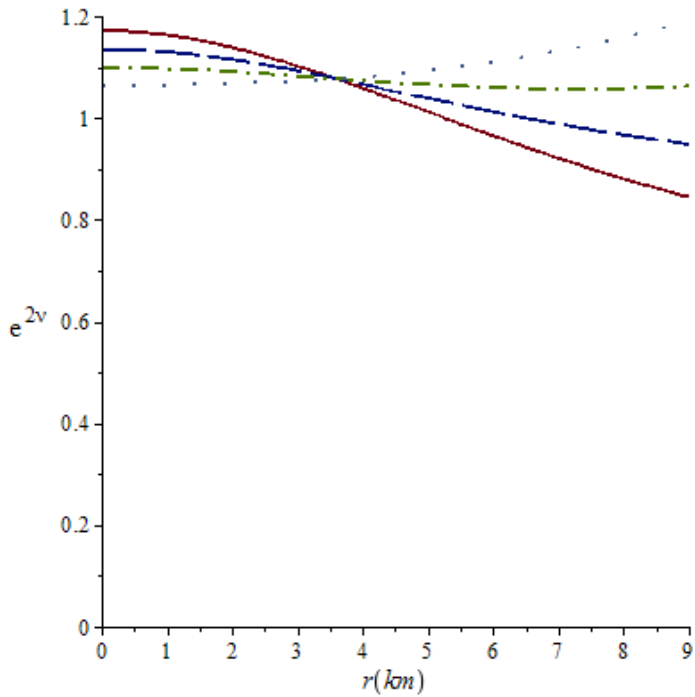

**Figure 8.** $e^{2\nu}$ against radial coordinate for ω=-1 (solid line); ω=-0.8 (long-dash line): ω=-0.6 (dashdot line); ω=-0.4 (spacedot line). In all the cases a=0.0064, α=2, c=1

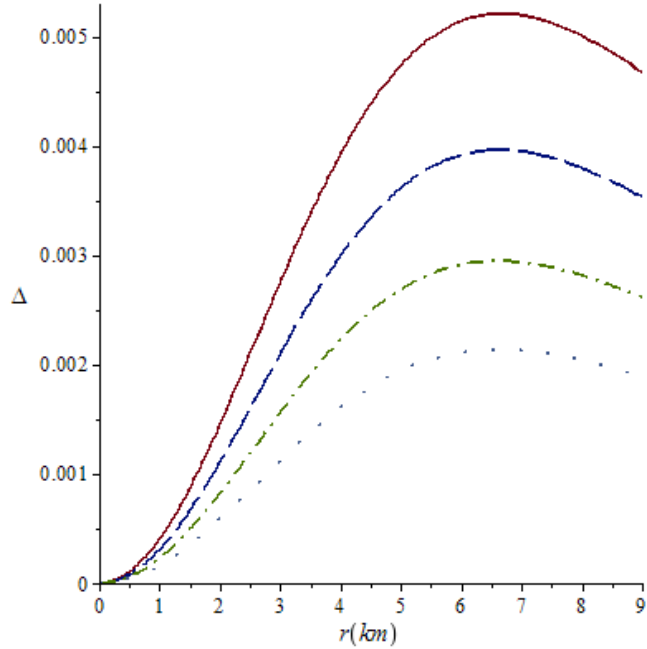

**Figure 9**. Δ against radial coordinate for ω=-1 (solid line); ω=-0.8 (long-dash line): ω=-0.6 (dashdot line); ω=-0.4 (spacedot line). In all the cases a=0.0064, α=2, c=1

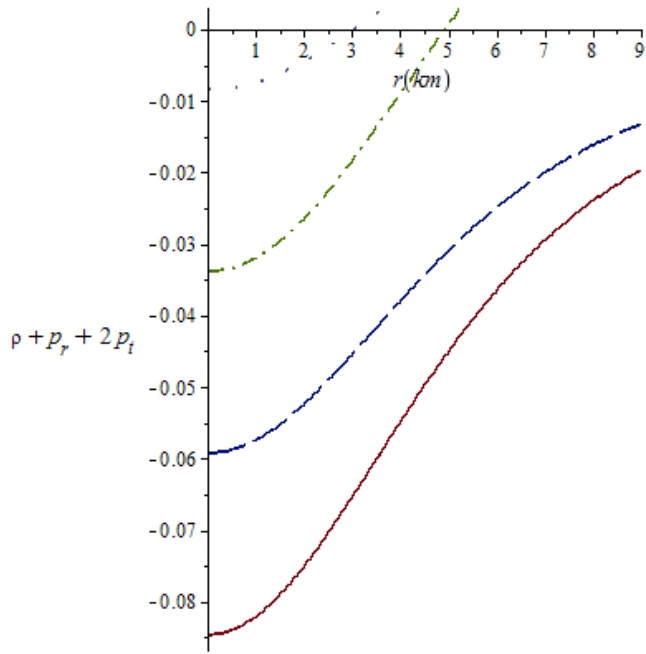

**Figure 10.** SEC against radial coordinate for ω=-1 (solid line); ω=-0.8 (long-dash line): ω=-0.6 (dashdot line); ω=-0.4 (spacedot line). In all the cases a=0.0064, α=2, c=1.

In figure 8, the metric potential $e^{2\nu}$ also is continuous, takes finite values and well behaved in the stellar interior for different values of *ω* and *α*=2 and in the Figure 9, the anisotropy is zero at the center *r=0* and its value decreases when *ω* increases. For all *ω* values considered, the violation of the strong energy condition is satisfied as noted in Figure 10. Further physical analysis has been provided in the conclusions of this paper.

Above results show that while metric potential, $e^{2\nu}$ continuously increases inside dark energy star modelled here. Energy density, $\rho$ seems to decrease through the interior of dark energy star, for the Figure 1 and anisotropic factor, Δ typically vanishes at the stellar center, according Figures 6 and 9. In addition, in Figure 7 strong energy condition (SEC) is violated for all *α* values considered when ω=-1, , and all *ω* values considered for the Figure 10. These aspects will mean typically quantum critical interactions with ordinary particles morphing into heavy non-relativistic-like particles, for example, quarks and gluons inside nucleons [11, 70] may not occur inside dark energy stars. However, the metric potential increasing inside dark energy stars will point to high zero-point energy gradient that have been modelled computed by ansatz general solutions of Schrödinger Helmholtz Hamiltonian quantum mechanics equations of vacuum quanta formalism [71,72]. However, questions of superfluids creating, for example microblackholes, that will probably have capability to generate quarks, antiquarks, as well as gluons out of vacuum quanta, with quasiperiodic oscillations, as also "Gravitybrane" or "Weakbrane" warped along the fifth dimension, having observables through gravitational wave astronomy measurements, alongside Hubble tension measurements remain open [73-79]. Observational evidence of the accelerated expansion of universe, considering quintessence dark energy field in the stellar model, typically free from central singularity [77] and a constant Early Dark Energy (EDE) derived model component, that takes on to account contributing a fraction ∼ 10% of the energy density of the universe and diluting as or faster than radiation has been able to provide a simple resolution to the Hubble tension measurements [7]. These advances will then further phenomenological emergent dark energy model where dark energy has no effective presence in the past and emerges at later times [79].

## 7. Conclusions

In this paper, we have presented the Gauss-Bonnet action in a n-dimensional space-time model with a modified version of metric potential proposed by Korkina-Orlyanskii.. A graphical analysis shows that the radial pressure, metric functions, energy density, mass function and anisotropy are regular at the origin and well behaved in the interior. The new solutions match smoothly with the exterior of the Einstein –Gauss-Bonnet- Schwarzschild at the boundary *r=R*, because matter variables and the gravitational potentials of this work are consistent with the physical analysis of these stars. It is expected that the results of this study can contribute to modeling of relativistic compact objects and configurations with anisotropic matter distribution. The *F(R)* Scalar-Gauss–Bonnet gravity equation of state and its modified version shall produce models associated to observed stellar objects and it makes our future works to further study in the upcoming papers.

The 5-D anisotropic metric geometry has been used to reproduce the modified Einstein equation. We have already told that the action is coupled with the Gauss-Bonnet lovelock reconstruction where the Gauss-Bonnet Lagrangian is coupled with Einstein action. This coupling is dependent upon the coupling coefficient α which later controls the evolution of all kind of variables useful in our calculation (Radial pressure, transverse pressure, anisotropy, energy density, density gradient, metric potentials and also the strong energy conditions). The pressure and density is discussed with the earlier assumed coordinate dependent function of $Z(x)$. But we found considerable changes when we change the coupling constant. That's why this coefficient has been renamed as a evolution controlling parameter in our calculation. The values of this parameter *α* has been changed between 0 to 6.

We have not used any direct dark energy models in our calculations instead we used the conditions for it. The dark energy dominated universe specially Quintessence universe belongs to the EOS parameter condition *-1 ≤ ω ≤ -1/3* .We know that the most established contradiction free dark energy model is quintessence model. When the EOS parameter has *ω < -1* , universe will face phantom universe and for *ω > -1/3* universe will feel Dark matter dominated (Hot Dark matter) phase. We have just discussed the Quintessence part by changing the EOS parameter between *-1* to *-1/3*. Late time acceleration, negativity of pressure, cosmic inflation can be well established with Quintessence model. The dynamics of stellar bodies, space time geometry will have effective modifications after using this dark energy condition. That's why we found differentiation in the results for different values of EOS parameters.

The application of coupling parameter and EOS parameter make considerable changes in the results of our calculations. Although the nature of the results remains conserved, we have found that the strong energy condition doesn't satisfy with the radial-transverse pressure and density modified with GB gravity into Quintessence region. From figure 1 to 7 we have provided the evolution of the variables in our calculations with changing the coupling parameter in the boundary mentioned earlier. We have observed that the values of those variables increase with increasing the value of coupling parameter. The values for different coupling coefficient coincides at large radial distance. This strictly says that for large distance the geometry effect of the GB modification becomes negligible. This matches the Einstein weak gravity approximation. The radial pressure has been found to be negative. This resolves the negative pressure problems in accelerated cosmology. This negativity has been found in absence of any direct application of Dark energy alternatives of gravity. The quintessence strictly provides acceleration and our results perfectly matched with this. The anisotropy initially increases and then decreases with distance. That mean we may conclude at the boundary where the anisotropy become maximum, should have some stellar body boundary which brings some kind of non-differentiable horizon in our system. On the other-hand the

SEC curves evolved such that at the large distance the universe should face deceleration. Thus, from the graphs of anisotropy and SEC we may conclude that the sudden change in anisotropy scheme was caused due to the change in model expansion from accelerating phase to decelerating phase at large distance. From figure 8 to 10 we see that the graphs decrease with increase the EOS parameter from *-1* to *-1/3*. The nature of the graphs still remains same as earlier.

Modified Einstein-Maxwell field equations with the Gauss-Bonnet coupling constant, applying modified form of gravitational potential proposed by Korkina-Orlyanskii have been transformed to algebraic equations as suggested by astrophysicists Durgapal and Bannerji [68] and Singh et al.[69] in order to obtain smooth physically meaningful solutions. Hence in this paper, we have generated a new class of solutions which represents a potential model of dark energy stars whose equation of state is $p_r = \omega\rho$ and with anisotropic matter distribution, within the framework of Einstein-Gauss-Bonnet (EGB) gravity.

Results highlight meaningfully useful output values of continuously increasing metric potential, decreasing energy density, and typically vanishing anisotropy factor inside the dark energy stars going towards center. Important results point to violation of strong energy conditions within the dark energy stars.

Physical analysis of these equations gave insight on the workings of these equations that may be interpreted to have non-occurrence of quantum critical interactions with ordinary particles morphing into heavy non-relativistic-like particles such as quarks and gluons inside nucleons. However, the metric potential increasing inside dark energy stars will point to high zero-point energy gradient that have been modelled by Iyer and Markoulakis [71]. Superfluids creating microblackholes, that will probably have capability to generate quarks, antiquarks, as well as gluons out of vacuum quanta, tested by observables through gravitational wave astronomy measurements, alongside Hubble tension measurements have open questions to answer. These solutions that we derived have generality to extend beyond and can be used in the development of the dark energy stars models, satisfying all physical acceptability conditions, but where the typical causality condition and strong energy condition cannot be satisfied.

Formalism advanced here has already generality to model further dark energy creation of stable geodesics within galactic star systems sustaining live cosmos, via originating supernovae stars evolving out dark energy stars.